\documentclass[aps,prl,showpacs,twocolumn]{revtex4}
\begin{document}
\def\a{\alpha}
\def\b{\beta}
\title{Anomalies of weakened decoherence criteria for quantum histories}
\author{Lajos Di\'osi}
\email{diosi@rmki.kfki.hu}
\homepage{www.rmki.kfki.hu/~diosi}
\affiliation{
Research Institute for Particle and Nuclear Physics\\
H-1525 Budapest 114, POB 49, Hungary}
\date{\today}

\begin{abstract}
The theory of decoherent histories is checked for the requirement of 
statistical independence of subsystems. Strikingly, this is satisfied
only when the decoherence functional is diagonal in both its real
{\it and\/} imaginary parts. In particular, the condition of
consistency (or weak decoherence) required for the assignment of
probabilities appears to be ruled out. The same conclusion is 
obtained independently, by claiming a plausible dynamical robustness 
of decoherent histories. 
\end{abstract}


\maketitle

{\it Introduction.}
Deriving testable statistical predictions from a given quantum state
$\rho$ is possible via decoherent histories (DH's)
\cite{Gri84,GMH90,Omn92,DowHal92,Hal94}, without
invoking von Neumann's concept \cite{Neu32} of quantum measurement 
and state reduction.
The definitive element of DH theory is a minimum mathematical
condition for the consistency of the probabilities $p_\a$ assigned to 
the histories $\a$. There has been a consensus that postulating the 
so-called {\it weak} decoherence condition:
\begin{equation}\label{ReDdiag}
{\rm Re} D(\a^\prime,\a)=0~,~~~~{\rm for~all}~\a\neq\a^\prime~~,
\end{equation}
for the decoherence functional $D(\a,\a^\prime)$ will assure
consistent probabilities. Term `weak' means we do not require the
diagonality of ${\rm Im}D(\a,\a^\prime)$.
Yet, this Letter presents two evidences
indicating that such weakened decoherence conditions are problematic.
The first evidence follows from trivial combination of statistically
independent subsystems. The second one follows from trivial modification
of the Hamiltonian dynamics. 

{\it Histories, decoherence.}
To define a {\it history}, one introduces a sequence of binary events 
for a succession of instances 
$t_1\langle t_2\langle\dots\langle t_n$:
\begin{equation}\label{Paks}
P_{\a_1}(t_1),P_{\a_2}(t_2),\dots,P_{\a_n}(t_n)~~.
\end{equation}
For each $k=1,2,\dots,n$, the $\{P_{\a_k}(t_k)\}$ are various complete 
sets of orthogonal projectors:
\begin{equation}\label{Pak}
\sum_{\a_k}P_{\a_k}(t_k)=I,~~~
P_{\a_k}(t_k)P_{\a_k^\prime}(t_k)=\delta_{\a_k\a_k^\prime}P_{\a_k}(t_k)~~.
\end{equation}
The history confining the events (\ref{Paks})
will be labeled by $\a=(\a_1,\dots,\a_n)$. All histories 
must be {\it consistent} in the sense that we can assign probabilities
to them. Introducing the time-ordered class operators
\begin{equation}\label{C}
C_\a=P_{\a_n}(t_n)\dots P_{\a_1}(t_1)~~,
\end{equation}
the following probability distribution is postulated:
\begin{equation}\label{pCC}
p_\a=\langle C_\a^\dagger C_\a\rangle_\rho~~,
\end{equation}
where $<>_\rho$ stands for expectation values in state $\rho$.
Concretely, consistency means the usual additivity of probabilities,
see Refs.\cite{GMH90,Omn92,DowHal92,Hal94} for standard explanation.
If, in particular, we bunch two different histories 
$\a$ and $\a^\prime$ into a coarse-grained one $\bar\a$:
\begin{equation}\label{abar}
{\bar C}_{\bar\a}=C_\a+C_{\a^\prime}~~,
\end{equation}
then the sum rule
\begin{equation}\label{sum}
{\bar p}_{\bar\a}=p_\a+p_{\a^\prime}
\end{equation}
must be satisfied for 
${\bar p}_{\bar\a}
=\langle {\bar C}_{\bar\a}^\dagger {\bar C}_{\bar\a} \rangle_\rho$. 
To guarantee this, one introduces \cite{GMH90} the 
{\it decoherence functional\/}:
\begin{equation}\label{D}
D(\a^\prime,\a)=\langle C_{\a^\prime}^\dagger C_\a \rangle_\rho~~.
\end{equation}
If $D(\a^\prime,\a)$ is diagonal in its double argument:
\begin{equation}\label{Ddiag}
D(\a^\prime,\a)=0~,~~~~{\rm for~all}~\a\neq\a^\prime~~,
\end{equation}
the consistency (\ref{sum}) of the probabilities (\ref{pCC}) is 
guaranteed by the absence of interference terms between
the contribution of the two histories $\a$ and $\a^\prime$. 
This is why Eq.(\ref{Ddiag}) is called {\it decoherence condition}. 
But it is not a necessary condition. The same interference terms 
cancel if we require the {\it weak} decoherence condition
({\ref{ReDdiag}) instead of the stronger condition (\ref{Ddiag}).

While weak decoherence was considered a kind of sufficient and 
necessary condition of consistency, Goldstein and Page \cite{GolPag94} 
suggested a radical generalization of DH's. In their theory of {\it 
linearly positive histories} no constraint is postulated on the 
decoherence functional (\ref{D}) and Eq.(\ref{pCC}) does not assign 
probabilities. A new equation, linear in $C_\a$, does:
\begin{equation}\label{RepC}
p_\a={\rm Re}\langle C_\a \rangle_\rho~~.
\end{equation}
The only postulated constraint is the natural one: 
\begin{equation}\label{ReCgeq0}
{\rm Re}\langle C_\a \rangle_\rho\geq 0~,~~~~{\rm for~all}~\a
\end{equation}
simply for the sake of non-negativity of probabilities 
$p_\a$ (\ref{RepC}). The consistency (\ref{sum}) of the probability 
assignment (\ref{ReCgeq0}) is guaranteed by construction. This concept 
represents further loosening weak decoherence. As shown in 
Ref.\cite{GolPag94}, weak DH's form a subset of linearly positive 
histories.

Surprisingly, neither the weak decoherent nor
the linear positive histories have so far been checked against common
tests such as system composition or trivial dynamic perturbations.
This Letter shows that these tests indicate serious inconsistencies
within the concept of weak decoherent and linearly positive histories.
 
{\it Test of composition.} 
Assume two statistically independent quantum systems $A$ and $B$ with 
states $\rho^A,\rho^B$, respectively. Let us assume that the class 
operators $C_\a^A,C_\b^B$ [c.f. Eq.(\ref{C})] generate consistent 
histories for $A$ and $B$ respectively. In ordinary quantum theory a 
trivial composition of two statistically independent subsystems is 
always possible. In our case the composite system's density operator 
is the direct product $\rho^A\otimes\rho^B$. It is plausible to expect 
that the operators $\{C_\a^A\otimes C_\b^B\}$ will generate DH's 
for the composite system. This latter's decoherence functional 
factorizes, in obvious notations it reads:
\begin{equation}\label{DAB}
D^{AB}(\a^\prime\b^\prime,\a\b)=D^A(\a^\prime,\a)D^B(\b^\prime,\b)~~.
\end{equation}
It is easy to see that the weak decoherence condition (\ref{ReDdiag}) 
for the statistically independent subsystems A and B does not imply the 
fulfillment of the same condition for the composite system. Using weak 
decoherence criterium, it may thus happen that we have DH's in 
subsystem $A$ and DH's in subsystem $B$ while the composition of those 
DH's are, contrary to our expectations, not DH's. This anomaly follows 
from mere composition of the two subsystems, without any correlation or 
interaction between them. Exactly the same anomaly appears in the theory 
of linearly positive histories. The reason is that the fulfillment of 
the positivity condition (\ref{ReCgeq0}) in the statistically 
independent subsystems can not imply its fulfillment in the composite 
system. 

{\it Test of dynamical stability.}
A given set of DH's may not persist if we alter the 
dynamics of the system. There are, nonetheless, situations when we
expect them to persist. Rather than pursuing the general case, 
let us consider the simplest one. Assume we switch on a sudden external
potential at $t_k+0$ (for a single $k$ between $1$ and $n$) if the 
binary variable $P_{\a_k}(t_k)$ at time $t_k$ takes value $1$. 
It it takes $0$, we do not alter the original dynamics. Let the 
corresponding interaction Hamiltonian be: 
\begin{equation}\label{dH}
\delta H(t)=\delta(t-t_k-0)
        \sum_{\a_k}\lambda_{\a_k}P_{\a_k}(t_k)~~,
\end{equation}
where the $\lambda_{\a_k}$ are real coupling constants. 
This Hamiltonian does not introduce coupling (coherence) between 
histories. We expect that consistency of histories is robust against it.
Fortunately, an analytic treatment is possible.
Under the perturbation (\ref{dH}), the time-ordered product (\ref{C}) 
of Heisenberg-operators changes in the following way:
\begin{equation}\label{CdH}
C_\a\rightarrow e^{-i\lambda_{\a_k}}U^\dagger(t_k)C_\a~~,
\end{equation} 
where $U(t_k)$ is the unitary transformation
\begin{equation}
U(t_k)=\exp\left(-i\sum_{\a_k}\lambda_{\a_k}(t_k)P_{\a_k}(t_k)\right)
\end{equation}
caused by the Hamiltonian (\ref{dH}) at time $t_k+0$.
By virtue of Eq.(\ref{CdH}), the decoherence functional (\ref{D}) 
changes as follows:
\begin{equation}
D(\a^\prime,\a)
\rightarrow e^{i(\lambda_{\a_k^\prime}-\lambda_{\a_k})}D(\a^\prime,\a)~~.
\end{equation}
It is now seen that the original decoherence criterium (\ref{Ddiag}) is
preserved whereas the weakened one (\ref{ReDdiag}) is not. Therefore, 
the DH's will loose their robustness against the trivial 
external fields (\ref{dH}) if we use the loosened (weak) decoherence 
condition. 

The same anomaly comes about for linearly positive histories.
Satisfying the positivity condition (\ref{ReCgeq0}) for the
unperturbed class operator $C$ can not assure the positivity condition 
after perturbation. Under perturbation (\ref{dH}), the probabilities 
(\ref{RepC}) would change as follows:
\begin{equation}
p_\a={\rm Re}\langle C_\a\rangle_\rho\rightarrow
     {\rm Re}e^{-i\lambda_{\a_k}}\langle U^\dagger(t_k)
     C_\a\rangle_\rho~~.
\end{equation}
Obviously, the preservation of positivity is not guaranteed. Linearly
positive histories may become lost under the influence of the trivial
external field, which is contrary to our expectations.

{\it Conclusion.}
We have argued that the complex decoherence functional must be diagonal.
The diagonality of its real part in itself is insufficient when 
checked for trivial composition of statistically independent subsystems.
This anomaly seems to exclude the so-called weak decoherence condition 
and urges to retain the stronger one.
The linearly positive histories show the same anomaly in composite 
systems. The survival of decoherent as well as of linearly positive 
histories has also been tested under trivial perturbation of the 
Hamiltonian. Expected survival has only been obtained at the stronger 
decoherence condition. 

The system composition evidence is likely to be an ultimate criticism. 
Any excuse should obviously question our standard notion of statistical
independence. Once Bell-inequalities changed our standard notion of 
statistical `dependence', without altering the more basic 
notion of statistical independence. This latter would be hard
to challenge for usual closed dynamic systems where DH theory had
originally been applied. 
The dynamical stability evidence is perhaps less convincing since
the claim of stability might need further theoretical support.
Finally we emphasize that {\it consistency} of histories had 
traditionally been restricted to the request of additivity of 
probabilities in coarse-grained histories. Consistency in composition 
or in perturbation had not been targeted apart from an early work 
\cite{Dio94} of the present author.

I thank Sheldon Goldstein for his contemporary remarks in 1994 on 
Ref.\cite{Dio94} as well as am I grateful to Jonathan Halliwell
for recent invaluable suggestions and for encouraging publication.


\begin{thebibliography}{99}
\bibitem{Gri84} R.B. Griffiths, J. Stat. Phys. {\bf 36} 219 (1984);
Phys. Rev. Lett. {\bf 70}, 2201 (1993).
\bibitem{GMH90} M. Gell-Mann and J.B. Hartle, in {\it Complexity,
Entropy and the Physics of Information}, Santa Fe Institute Studies
in the Science of Complexity, Vol. VIII edited by W. Zurek
(Addison-Wesley, Redwood City, CA, 1990); Phys. Rev. {\bf D 47}, 
3345 (1993).
\bibitem{Omn92} R. Omn\`es, Rev. Mod. Phys. {\bf 64}, 339 (1992).
\bibitem{DowHal92} H.F. Dowker and J.J. Halliwell, Phys. Rev. 
{\bf D 46}, 1580 (1992).
\bibitem{Hal94} J.J.Halliwell, in {\it Stochastic Evolution of Quantum
States in Open Systems and Measurement Processes}, eds. L. Di\'osi and 
B. Luk\'acs (World Scientific, Singapore, 1994), gr-qc/9308005; 
Phys. Rev. Lett. {\bf 83}, 2481 (1999).
\bibitem{DodHal03} P.J. Dodd and J.J. Halliwell, Phys. Rev. {\bf D67},
105018 (2003).
\bibitem{Neu32} J. von Neumann, {\it Mathematical Foundations of 
Quantum Mechanics}, (Princeton Univ. Press, Princeton, 1955).
\bibitem{GolPag94} S.Goldstein and D.N.Page, Phys.Rev.Lett. {\bf 74}, 
3715 (1995).
\bibitem{Dio94} L. Di\'osi, gr-qc/9409017.
\end{thebibliography}
\end{document}